\documentclass[final]{IEEEtran}
\usepackage[update,prepend]{epstopdf}

\usepackage{graphics}
\usepackage{multirow}
\usepackage{tikz}
\usepackage{bbm} 
\usepackage{pdfpages}
\usepackage{subfig}
\usepackage{comment}

\usepackage{setspace}
\usepackage{graphicx}
\usepackage{algorithm,algorithmic}
\usepackage{multicol}

\usepackage[justification=centering]{caption}
\usepackage{textcomp}
\usepackage{psfrag}
\usepackage{arydshln}
\usepackage{url}
\usepackage{soul}
\usepackage{graphicx,color}
\usepackage[nolist]{acronym}

\usepackage{mathtools,lipsum}
\usepackage{cuted}
\usepackage{amsmath}
\usepackage{graphicx}
\usepackage[colorlinks,citecolor=red, linkcolor=blue, anchorcolor=green]{hyperref}

\def\nb0{{\mathbf{0}}}
\def\nb1{{\mathbf{1}}}

\allowdisplaybreaks 

\begin{document}
\graphicspath{{./Figures/}{./figures/}}
	\begin{acronym}

\acro{5G-NR}{5G New Radio}
\acro{3GPP}{3rd Generation Partnership Project}
\acro{ABS}{aerial base station}
\acro{AC}{address coding}
\acro{ACF}{autocorrelation function}
\acro{ACR}{autocorrelation receiver}
\acro{ADC}{analog-to-digital converter}
\acrodef{aic}[AIC]{Analog-to-Information Converter}     
\acro{AIC}[AIC]{Akaike information criterion}
\acro{aric}[ARIC]{asymmetric restricted isometry constant}
\acro{arip}[ARIP]{asymmetric restricted isometry property}

\acro{ARQ}{Automatic Repeat Request}
\acro{AUB}{asymptotic union bound}
\acrodef{awgn}[AWGN]{Additive White Gaussian Noise}     
\acro{AWGN}{additive white Gaussian noise}

\acro{APSK}[PSK]{asymmetric PSK} 

\acro{waric}[AWRICs]{asymmetric weak restricted isometry constants}
\acro{warip}[AWRIP]{asymmetric weak restricted isometry property}
\acro{BCH}{Bose, Chaudhuri, and Hocquenghem}        
\acro{BCHC}[BCHSC]{BCH based source coding}
\acro{BEP}{bit error probability}
\acro{BFC}{block fading channel}
\acro{BG}[BG]{Bernoulli-Gaussian}
\acro{BGG}{Bernoulli-Generalized Gaussian}
\acro{BPAM}{binary pulse amplitude modulation}
\acro{BPDN}{Basis Pursuit Denoising}
\acro{BPPM}{binary pulse position modulation}
\acro{BPSK}{Binary Phase Shift Keying}
\acro{BPZF}{bandpass zonal filter}
\acro{BSC}{binary symmetric channels}              
\acro{BU}[BU]{Bernoulli-uniform}
\acro{BER}{bit error rate}
\acro{BS}{base station}
\acro{BW}{BandWidth}
\acro{BLLL}{ binary log-linear learning }

\acro{CP}{Cyclic Prefix}
\acrodef{cdf}[CDF]{cumulative distribution function}   
\acro{CDF}{Cumulative Distribution Function}
\acrodef{c.d.f.}[CDF]{cumulative distribution function}
\acro{CCDF}{complementary cumulative distribution function}
\acrodef{ccdf}[CCDF]{complementary CDF}               
\acrodef{c.c.d.f.}[CCDF]{complementary cumulative distribution function}
\acro{CD}{cooperative diversity}

\acro{CDMA}{Code Division Multiple Access}
\acro{ch.f.}{characteristic function}
\acro{CIR}{channel impulse response}
\acro{cosamp}[CoSaMP]{compressive sampling matching pursuit}
\acro{CR}{cognitive radio}
\acro{cs}[CS]{compressed sensing}                   
\acrodef{cscapital}[CS]{Compressed sensing} 
\acrodef{CS}[CS]{compressed sensing}
\acro{CSI}{channel state information}
\acro{CCSDS}{consultative committee for space data systems}
\acro{CC}{convolutional coding}
\acro{Covid19}[COVID-19]{Coronavirus disease}

\acro{DAA}{detect and avoid}
\acro{DAB}{digital audio broadcasting}
\acro{DCT}{discrete cosine transform}
\acro{dft}[DFT]{discrete Fourier transform}
\acro{DR}{distortion-rate}
\acro{DS}{direct sequence}
\acro{DS-SS}{direct-sequence spread-spectrum}
\acro{DTR}{differential transmitted-reference}
\acro{DVB-H}{digital video broadcasting\,--\,handheld}
\acro{DVB-T}{digital video broadcasting\,--\,terrestrial}
\acro{DL}{DownLink}
\acro{DSSS}{Direct Sequence Spread Spectrum}
\acro{DFT-s-OFDM}{Discrete Fourier Transform-spread-Orthogonal Frequency Division Multiplexing}
\acro{DAS}{Distributed Antenna System}
\acro{DNA}{DeoxyriboNucleic Acid}

\acro{EC}{European Commission}
\acro{EED}[EED]{exact eigenvalues distribution}
\acro{EIRP}{Equivalent Isotropically Radiated Power}
\acro{ELP}{equivalent low-pass}
\acro{eMBB}{Enhanced Mobile Broadband}
\acro{EMF}{ElectroMagnetic Field}
\acro{EU}{European union}
\acro{EI}{Exposure Index}
\acro{eICIC}{enhanced Inter-Cell Interference Coordination}

\acro{FC}[FC]{fusion center}
\acro{FCC}{Federal Communications Commission}
\acro{FEC}{forward error correction}
\acro{FFT}{fast Fourier transform}
\acro{FH}{frequency-hopping}
\acro{FH-SS}{frequency-hopping spread-spectrum}
\acrodef{FS}{Frame synchronization}
\acro{FSsmall}[FS]{frame synchronization}  
\acro{FDMA}{Frequency Division Multiple Access}

\acro{GA}{Gaussian approximation}
\acro{GF}{Galois field }
\acro{GG}{Generalized-Gaussian}
\acro{GIC}[GIC]{generalized information criterion}
\acro{GLRT}{generalized likelihood ratio test}
\acro{GPS}{Global Positioning System}
\acro{GMSK}{Gaussian Minimum Shift Keying}
\acro{GSMA}{Global System for Mobile communications Association}
\acro{GS}{ground station}
\acro{GMG}{ Grid-connected MicroGeneration}

\acro{HAP}{high altitude platform}
\acro{HetNet}{Heterogeneous network}

\acro{IDR}{information distortion-rate}
\acro{IFFT}{inverse fast Fourier transform}
\acro{iht}[IHT]{iterative hard thresholding}
\acro{i.i.d.}{independent, identically distributed}
\acro{IoT}{Internet of Things}                      
\acro{IR}{impulse radio}
\acro{lric}[LRIC]{lower restricted isometry constant}
\acro{lrict}[LRICt]{lower restricted isometry constant threshold}
\acro{ISI}{intersymbol interference}
\acro{ITU}{International Telecommunication Union}
\acro{ICNIRP}{International Commission on Non-Ionizing Radiation Protection}
\acro{IEEE}{Institute of Electrical and Electronics Engineers}
\acro{ICES}{IEEE international committee on electromagnetic safety}
\acro{IEC}{International Electrotechnical Commission}
\acro{IARC}{International Agency on Research on Cancer}
\acro{IS-95}{Interim Standard 95}

\acro{KPI}{Key Performance Indicator}

\acro{LEO}{low earth orbit}
\acro{LF}{likelihood function}
\acro{LLF}{log-likelihood function}
\acro{LLR}{log-likelihood ratio}
\acro{LLRT}{log-likelihood ratio test}
\acro{LoS}{Line-of-Sight}
\acro{LRT}{likelihood ratio test}
\acro{wlric}[LWRIC]{lower weak restricted isometry constant}
\acro{wlrict}[LWRICt]{LWRIC threshold}
\acro{LPWAN}{Low Power Wide Area Network}
\acro{LoRaWAN}{Low power long Range Wide Area Network}
\acro{NLoS}{Non-Line-of-Sight}
\acro{LiFi}[Li-Fi]{light-fidelity}
 \acro{LED}{light emitting diode}
 \acro{LABS}{LoS transmission with each ABS}
 \acro{NLABS}{NLoS transmission with each ABS}

\acro{MB}{multiband}
\acro{MC}{macro cell}
\acro{MDS}{mixed distributed source}
\acro{MF}{matched filter}
\acro{m.g.f.}{moment generating function}
\acro{MI}{mutual information}
\acro{MIMO}{Multiple-Input Multiple-Output}
\acro{MISO}{multiple-input single-output}
\acrodef{maxs}[MJSO]{maximum joint support cardinality}                       
\acro{ML}[ML]{maximum likelihood}
\acro{MMSE}{minimum mean-square error}
\acro{MMV}{multiple measurement vectors}
\acrodef{MOS}{model order selection}
\acro{M-PSK}[${M}$-PSK]{$M$-ary phase shift keying}                       
\acro{M-APSK}[${M}$-PSK]{$M$-ary asymmetric PSK} 
\acro{MP}{ multi-period}
\acro{MINLP}{mixed integer non-linear programming}

\acro{M-QAM}[$M$-QAM]{$M$-ary quadrature amplitude modulation}
\acro{MRC}{maximal ratio combiner}                  
\acro{maxs}[MSO]{maximum sparsity order}                                      
\acro{M2M}{Machine-to-Machine}                                                
\acro{MUI}{multi-user interference}
\acro{mMTC}{massive Machine Type Communications}      
\acro{mm-Wave}{millimeter-wave}
\acro{MP}{mobile phone}
\acro{MPE}{maximum permissible exposure}
\acro{MAC}{media access control}
\acro{NB}{narrowband}
\acro{NBI}{narrowband interference}
\acro{NLA}{nonlinear sparse approximation}
\acro{NLOS}{Non-Line of Sight}
\acro{NTIA}{National Telecommunications and Information Administration}
\acro{NTP}{National Toxicology Program}
\acro{NHS}{National Health Service}

\acro{LOS}{Line of Sight}

\acro{OC}{optimum combining}                             
\acro{OC}{optimum combining}
\acro{ODE}{operational distortion-energy}
\acro{ODR}{operational distortion-rate}
\acro{OFDM}{Orthogonal Frequency-Division Multiplexing}
\acro{omp}[OMP]{orthogonal matching pursuit}
\acro{OSMP}[OSMP]{orthogonal subspace matching pursuit}
\acro{OQAM}{offset quadrature amplitude modulation}
\acro{OQPSK}{offset QPSK}
\acro{OFDMA}{Orthogonal Frequency-division Multiple Access}
\acro{OPEX}{Operating Expenditures}
\acro{OQPSK/PM}{OQPSK with phase modulation}

\acro{PAM}{pulse amplitude modulation}
\acro{PAR}{peak-to-average ratio}
\acrodef{pdf}[PDF]{probability density function}                      
\acro{PDF}{probability density function}
\acrodef{p.d.f.}[PDF]{probability distribution function}
\acro{PDP}{power dispersion profile}
\acro{PMF}{probability mass function}                             
\acrodef{p.m.f.}[PMF]{probability mass function}
\acro{PN}{pseudo-noise}
\acro{PPM}{pulse position modulation}
\acro{PRake}{Partial Rake}
\acro{PSD}{power spectral density}
\acro{PSEP}{pairwise synchronization error probability}
\acro{PSK}{phase shift keying}
\acro{PD}{power density}
\acro{8-PSK}[$8$-PSK]{$8$-phase shift keying}
\acro{PPP}{Poisson point process}
\acro{PCP}{Poisson cluster process}
 
\acro{FSK}{Frequency Shift Keying}

\acro{QAM}{Quadrature Amplitude Modulation}
\acro{QPSK}{Quadrature Phase Shift Keying}
\acro{OQPSK/PM}{OQPSK with phase modulator }

\acro{RD}[RD]{raw data}
\acro{RDL}{"random data limit"}
\acro{ric}[RIC]{restricted isometry constant}
\acro{rict}[RICt]{restricted isometry constant threshold}
\acro{rip}[RIP]{restricted isometry property}
\acro{ROC}{receiver operating characteristic}
\acro{rq}[RQ]{Raleigh quotient}
\acro{RS}[RS]{Reed-Solomon}
\acro{RSC}[RSSC]{RS based source coding}
\acro{r.v.}{random variable}                               
\acro{R.V.}{random vector}
\acro{RMS}{root mean square}
\acro{RFR}{radiofrequency radiation}
\acro{RIS}{Reconfigurable Intelligent Surface}
\acro{RNA}{RiboNucleic Acid}
\acro{RRM}{Radio Resource Management}
\acro{RUE}{reference user equipments}
\acro{RAT}{radio access technology}
\acro{RB}{resource block}

\acro{SA}[SA-Music]{subspace-augmented MUSIC with OSMP}
\acro{SC}{small cell}
\acro{SCBSES}[SCBSES]{Source Compression Based Syndrome Encoding Scheme}
\acro{SCM}{sample covariance matrix}
\acro{SEP}{symbol error probability}
\acro{SG}[SG]{sparse-land Gaussian model}
\acro{SIMO}{single-input multiple-output}
\acro{SINR}{signal-to-interference plus noise ratio}
\acro{SIR}{signal-to-interference ratio}
\acro{SISO}{Single-Input Single-Output}
\acro{SMV}{single measurement vector}
\acro{SNR}[\textrm{SNR}]{signal-to-noise ratio} 
\acro{sp}[SP]{subspace pursuit}
\acro{SS}{spread spectrum}
\acro{SW}{sync word}
\acro{SAR}{specific absorption rate}
\acro{SSB}{synchronization signal block}
\acro{SR}{shrink and realign}

\acro{tUAV}{tethered Unmanned Aerial Vehicle}
\acro{TBS}{terrestrial base station}

\acro{uUAV}{untethered Unmanned Aerial Vehicle}
\acro{PDF}{probability density functions}

\acro{PL}{path-loss}

\acro{TH}{time-hopping}
\acro{ToA}{time-of-arrival}
\acro{TR}{transmitted-reference}
\acro{TW}{Tracy-Widom}
\acro{TWDT}{TW Distribution Tail}
\acro{TCM}{trellis coded modulation}
\acro{TDD}{Time-Division Duplexing}
\acro{TDMA}{Time Division Multiple Access}
\acro{Tx}{average transmit}

\acro{UAV}{Unmanned Aerial Vehicle}
\acro{uric}[URIC]{upper restricted isometry constant}
\acro{urict}[URICt]{upper restricted isometry constant threshold}
\acro{UWB}{ultrawide band}
\acro{UWBcap}[UWB]{Ultrawide band}   
\acro{URLLC}{Ultra Reliable Low Latency Communications}
         
\acro{wuric}[UWRIC]{upper weak restricted isometry constant}
\acro{wurict}[UWRICt]{UWRIC threshold}                
\acro{UE}{User Equipment}
\acro{UL}{UpLink}

\acro{WiM}[WiM]{weigh-in-motion}
\acro{WLAN}{wireless local area network}
\acro{wm}[WM]{Wishart matrix}                               
\acroplural{wm}[WM]{Wishart matrices}
\acro{WMAN}{wireless metropolitan area network}
\acro{WPAN}{wireless personal area network}
\acro{wric}[WRIC]{weak restricted isometry constant}
\acro{wrict}[WRICt]{weak restricted isometry constant thresholds}
\acro{wrip}[WRIP]{weak restricted isometry property}
\acro{WSN}{wireless sensor network}                        
\acro{WSS}{Wide-Sense Stationary}
\acro{WHO}{World Health Organization}
\acro{Wi-Fi}{Wireless Fidelity}

\acro{sss}[SpaSoSEnc]{sparse source syndrome encoding}

\acro{VLC}{Visible Light Communication}
\acro{VPN}{Virtual Private Network} 
\acro{RF}{Radio Frequency}
\acro{FSO}{Free Space Optics}
\acro{IoST}{Internet of Space Things}

\acro{GSM}{Global System for Mobile Communications}
\acro{2G}{Second-generation cellular network}
\acro{3G}{Third-generation cellular network}
\acro{4G}{Fourth-generation cellular network}
\acro{5G}{Fifth-generation cellular network}	
\acro{gNB}{next-generation Node-B Base Station}
\acro{NR}{New Radio}
\acro{UMTS}{Universal Mobile Telecommunications Service}
\acro{LTE}{Long Term Evolution}

\acro{QoS}{Quality of Service}
\end{acronym}

\newcommand{\SAR} {\mathrm{SAR}}
\newcommand{\WBSAR} {\mathrm{SAR}_{\mathsf{WB}}}
\newcommand{\gSAR} {\mathrm{SAR}_{10\si{\gram}}}
\newcommand{\Sab} {S_{\mathsf{ab}}}
\newcommand{\Eavg} {E_{\mathsf{avg}}}
\newcommand{\ft}{f_{\textsf{th}}}
\newcommand{\alphatf}{\alpha_{24}}

\title{
Hierarchical Edge Computing in SAGSIN:  \\ Multi-Layer Network Architecture and Multi-Level Information Processing
}

\author{
Jiajie Xu,~\IEEEmembership{Member,~IEEE,} Zhengying Lou, and Mohamed-Slim Alouini, {\em Fellow, IEEE}

\thanks{Jiajie Xu  and Mohamed-Slim Alouini are with the King Abdullah
University of Science and Technology, Thuwal 23955, Saudi Arabia. e-mail: \{jiajie.xu.1, slim.alouini\}@kaust.edu.sa. 

Z. Lou is with the Division of Engineering and the Wireless Center, New York University Abu Dhabi, PO Box 129188, Abu Dhabi, UAE, e-mails: zl6994@nyu.edu. This work was conducted while Z. Lou was a Ph.D. student at KAUST. 
}

\vspace{-6mm}
}

\maketitle
\thispagestyle{empty}
\pagestyle{empty}

\begin{abstract}
Maritime Internet of Things (IoT) deployments increasingly rely on the space--air--ground--sea integrated network (SAGSIN) to connect underwater sensors with terrestrial and space backbones. However, the heterogeneous links along this path, ranging from bandwidth-limited and energy-hungry underwater acoustic channels to high-capacity optical links above the sea, make the transport of massive raw sensing data costly in both latency and energy, and difficult to sustain for unattended, battery-powered nodes. This article presents an edge-computing paradigm for SAGSIN built on two coupled ideas: a Multi-Layer Network Architecture (MLNA) that organizes the underwater, surface, aerial, and ground/space tiers, and Multi-Level Information Processing (MLIP) that progressively refines data from raw measurements toward compact, event-level representations as they ascend the network. We characterize the computation, transmission, and storage energy at each tier and show how distributing refinement across layers trades local processing cost against transmission and storage savings. We then discuss how MLNA--MLIP reduces latency, strengthens data privacy, improves service reliability, and manages energy to prolong network lifetime. A case study on offshore monitoring quantifies the resulting lifetime gains and identifies the optimal processing depth. Open challenges and future directions are outlined.
\end{abstract}

\begin{IEEEkeywords}
Maritime IoT, SAGSIN, edge computing, multi-level information processing,   energy efficiency, network lifetime.
\end{IEEEkeywords}

\vspace{-0.2cm}

\section{Introduction}

The ocean covers more than 70\% of the Earth's surface, yet remains sparsely instrumented compared with terrestrial environments. Emerging maritime applications, environmental observation, seismic and tsunami early warning, and offshore resource exploration are driving large-scale deployments of maritime Internet of Things (IoT), in which underwater sensors must ultimately deliver their measurements to a terrestrial or space-based core network. The space--air--ground--sea integrated network (SAGSIN) has emerged as the natural substrate for this end-to-end connectivity \cite{10040542}, stitching together underwater acoustic links, surface relays, stratospheric platforms, and satellite constellations into a single multi-tier system \cite{11531198}.

Delivering raw sensing data across this entire path, however, is both slow and energy-intensive. Underwater acoustic links offer only tens of kbps of bandwidth, incur large propagation delays, and consume substantial energy per bit, with additional overhead spent on wake-up, synchronization, and handshaking before each transmission \cite{ref-uwsn}.
Above the sea surface, radio and optical links are faster but remain weather-dependent and still carry a non-trivial per-bit cost over long ranges. Meanwhile, underwater sensor nodes (USNs) are battery-powered, deployed in harsh environments, and almost impossible to recharge or service \cite{ref-uwsn}.
Streaming every raw measurement from the seabed to the core network is therefore unsustainable in terms of latency and energy and ultimately limits network lifetime.

{\color{black} Edge computing offers a way out by moving processing closer to where data are generated. Most existing efforts, however,  treat ``how much a measurement should be processed'' as a fixed, node-local decision.
For example, hierarchical edge computing partitions an application into predefined sensor-side and edge-side processing tasks \cite{8769942}, while underwater clustering schemes apply a fixed aggregation operation before forwarding data from cluster heads \cite{ref-offload}. 
Hierarchical filtering, compression, and feature extraction are established processing operations.
What is missing is a unified view that spans the full underwater-to-space path and explicitly decides \emph{to which level of refinement} data should be processed at \emph{which tier}, given the very different energy costs along the way.}

{\color{black} To address this gap, this article proposes a framework termed Multi-Layer Network Architecture and Multi-Level Information Processing (MLNA--MLIP), which couples information-refinement depth with processing placement across SAGSIN tiers.
First, we organize SAGSIN into a four-tier MLNA spanning underwater sensing, surface cross-medium transformation, aerial forwarding, and ground/space aggregation. Second, we introduce MLIP, a hierarchy of information levels $L_0 \rightarrow L_N$ along which data are progressively filtered, compressed, abstracted into features, and finally distilled into events.
Third, we characterize the three energy components, computation, transmission, and storage, that govern this refinement and expose the cross-layer trade-off between local processing cost and downstream transmission/storage savings.
Fourth, we discuss how MLNA--MLIP simultaneously reduces latency, enhances data privacy, improves service reliability, and manages energy to prolong network lifetime. Finally, a case study on offshore monitoring quantifies the lifetime gains and reveals an optimal processing depth that depends on the tier and link in use. We then discuss the remaining open challenges.}

\begin{figure*}[!t]
    \vspace{-3mm}
    \centering  \includegraphics[width=1\linewidth]{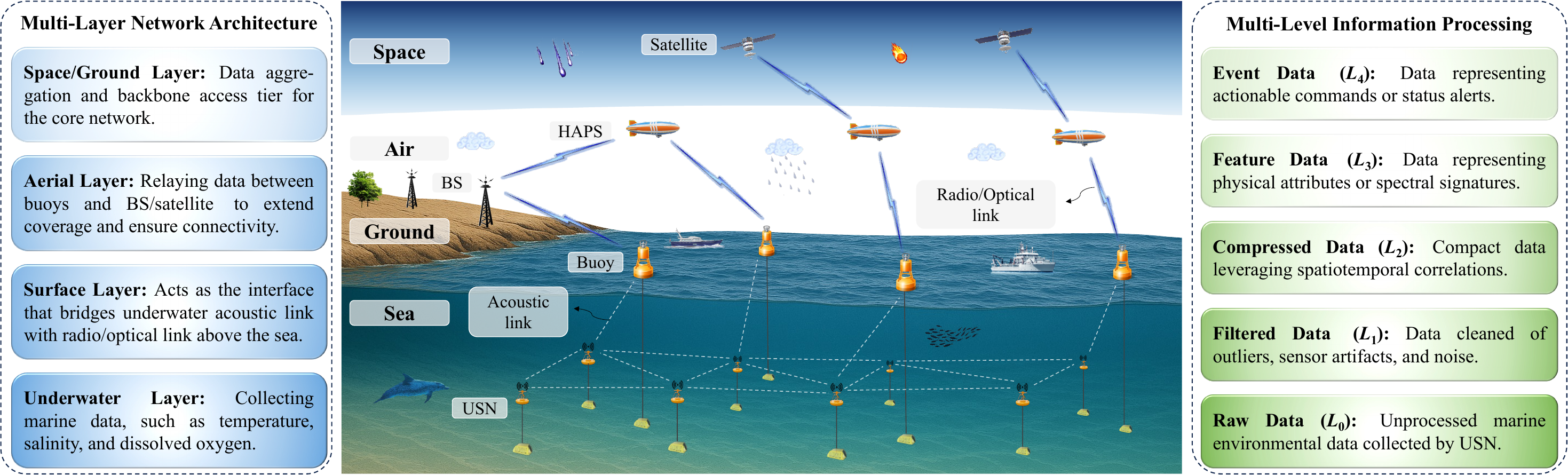}
    \caption{Illustration of MLNA-MLIP in SAGSIN.}
    \label{fig:overleap}
    \vspace{-3mm}
\end{figure*}

\section{Multi-Layer Network Architecture}

The MLNA--MLIP framework rests on two complementary pillars. The first, MLNA, organizes SAGSIN into four cooperative tiers---underwater, surface, aerial, and ground/space---each with distinct devices, energy budgets, and link characteristics. The second, MLIP, defines how sensing data are progressively refined from raw measurements into compact, event-level information as they traverse these tiers. Fig.~\ref{fig:overleap} illustrates how the two pillars operate jointly: maritime data are collected at the seabed, relayed upward across the four layers, and refined level by level along the way, so that each hop carries only the information its downstream link can afford. The following two sections detail the network architecture and the information-processing hierarchy in turn.

{\color{black}
To support large-scale maritime IoT applications, SAGSIN integrates four complementary tiers in which underwater nodes provide sensing and acoustic access, surface buoys perform cross-medium conversion, aerial relays extend coverage beyond terrestrial reach, and ground/space infrastructure provides data aggregation and backbone connectivity.}

\subsection{Underwater Layer: Data Collection}

The underwater layer is responsible for continuously sensing and collecting marine data, such as temperature, salinity, pressure, and dissolved oxygen.
This layer primarily consists of a large number of fixed underwater sensor nodes (USNs) anchored to the seabed in grid-based topologies for stable long-term monitoring.
In addition, autonomous underwater vehicles (AUVs) are employed as mobile complements to the fixed network, extending spatial coverage.

Fixed USNs are typically battery-powered and hard to maintain in the harsh marine environment.
A typical node is equipped with a lithium battery pack of 100 to 500 Wh, providing an operational lifetime of several months. To prolong the lifetime, energy-efficient strategies such as clustering and adaptive duty cycling are widely employed.
In addition, AUVs can operate for tens of hours on a single charge before returning to the surface for recharging.

Underwater communication mainly relies on acoustic links, which can span tens of kilometers, whereas radio and optical signals are limited to only tens of meters due to strong attenuation in seawater.
Despite their long range, acoustic channels provide limited bandwidth, typically between 1 and 10 kbps. Moreover, the required transmission power increases with communication distance because of geometric spreading and frequency-dependent absorption, reaching tens of watts for a multi-kilometer link.
Furthermore, acoustic links suffer from large propagation delays due to the low sound speed of 1.5 km/s.

\subsection{Surface Layer: Cross-Medium Transformation}

The surface layer acts as the cross-medium interface that bridges underwater acoustic communications with radio or optical links above the sea.
Two types of buoys are typically employed depending on the application and environmental conditions.
Among them, moored buoys are predominantly used in marine IoT networks, as they are anchored to the seabed using steel, providing stable positioning and reliable communication links for long-term observation and data relay.
In contrast, drifting buoys float freely with ocean currents and winds, offering greater flexibility and coverage over wide areas but with reduced positional stability.

Surface buoys are designed for long-term autonomous operation and are typically powered by renewable energy sources combined with onboard batteries.
Solar panels are the most common energy source, often complemented by wave or wind energy harvesters to ensure a continuous power supply under variable weather conditions.
The harvested energy is stored in battery packs with capacities ranging from a few hundred Wh to several kWh, depending on the buoy's size and operational requirements.
With hybrid power management and duty-cycling strategies, surface buoys can operate continuously for six months or longer without maintenance.

When operating near the coastline (within 20 km), buoys forward the collected data to base stations (BS) on the shore, whereas in offshore regions, the data are relayed to high-altitude platforms (HAPS) in the stratosphere.
Links between buoys and BS/HAPS typically operate in the sub-6 GHz band, ensuring all-weather availability above 99\% even under high humidity and aerosol concentration.
To further enhance capacity, free-space optical (FSO) links can be integrated as a complementary channel.
FSO provides Gbps-level throughput under clear-sky conditions but is sensitive to fog, clouds, and sea spray \cite{ref-fso}.

\subsection{Aerial Layer: Connectivity Bridge}

The Aerial Layer serves as a relay that bridges the surface layer and the space/ground layer to extend coverage and ensure connectivity.
It is composed of HAPS operating at altitudes of 17--22 km, where the atmosphere is stable and nearly free of clouds and turbulence \cite{ref-haps}.
Owing to this elevated position, the communication link between surface buoys and HAPS is minimally affected by sea-surface reflections or the Earth's curvature.
Unlike LEO satellites that move rapidly and require frequent handovers, HAPS remain quasi-stationary in the stratosphere.
A single HAPS can provide coverage with a radius of at least 100 km, enabling continuous connectivity over wide oceanic areas.

HAPS derive their energy primarily from solar power, utilizing high-efficiency photovoltaic arrays to harvest energy in the stratosphere.
The collected energy not only sustains daytime operations but also charges onboard batteries that provide power during nighttime.
The major sources of energy consumption include the propulsion system for station-keeping, the communication payload for signal processing and data transmission, and the thermal control unit for temperature stabilization under diurnal variations.
To enable long-term operation, aerostatic HAPS employ helium-based buoyant lift to reduce propulsion energy consumption while supporting payloads of hundreds of kilograms.

Depending on visibility and weather, the HAPS forwards maritime data either to ground BS or to LEO satellites, both of which interface with the core network.
The HAPS-BS link traverses the troposphere, so availability is dominated by rain and fog.
{\color{black} Sub-6 GHz provides all-weather robustness for control and fallback, whereas mmWave offers high capacity with narrow-beam arrays but requires adaptive coding and modulation, precise pointing, and rain-fade margins.}
Moreover, the optical link can be added as a complementary clear-sky channel, delivering multi-Gb/s.
In contrast, the HAPS--LEO link is essentially free-space above the troposphere.
{\color{black} Accordingly, mmWave and optical links can deliver high capacity when supported by rapid pointing-acquisition--tracking and precise beam control.
}

\subsection{Ground/Space Layer: Data Aggregation}

The ground/space layer functions as a data aggregation and backbone access tier for the core network.
The ground layer consists of fixed BS, typically installed on towers at heights of 10--50 m. 
{\color{black} Owing to the Earth's curvature, the elevation of the BS determines the line-of-sight coverage range, which typically extends up to 20--40 km, providing stable connectivity for nearshore maritime regions.}
In contrast, the space layer is composed of LEO satellites operating at altitudes of 400--2,000 km \cite{ref-ntn}.
LEO satellites move rapidly relative to Earth, traveling at velocities of 6--8 km/s and completing one revolution every 90--120 minutes. At these altitudes, each satellite covers a footprint of 1,000 km in radius and remains visible to a ground terminal for 5-10 minutes.
Continuous service is maintained through coordinated handovers within large constellations.

Both BSs and LEO satellites exhibit stable operational lifetimes exceeding 5 years.
Specifically, BS are powered by the electrical grid, ensuring a stable and continuous energy supply for transmission and control operations. The lifetime of BSs is mainly constrained by the aging of electrical and electronic components. Nevertheless, modular hardware design and scheduled maintenance can significantly extend their operational lifespan, often exceeding 20 years.
As for LEO satellites, they rely primarily on solar arrays, while lithium-ion batteries sustain operation during orbital eclipses.
Their energy system continuously balances power generation, storage, and consumption to support communication payloads, attitude control, and thermal regulation.
To maintain constellation performance and ensure continuous global coverage, aging satellites are periodically replaced.

{\color{black} The four layers impose different processing, communication, and control
burdens. The underwater layer must coordinate sensing and local refinement
under severe battery constraints and low-rate, high-latency acoustic links.
The surface layer performs cross-medium adaptation, buffering, and translation
between underwater acoustic and above-water radio or optical interfaces.
The aerial layer must jointly manage forwarding, link selection, and
weather-dependent connectivity under a variable energy supply. The
ground/space layer aggregates traffic while handling satellite mobility,
handover, service continuity, and cross-domain resource coordination.}
\section{Multi-Level Information Processing} \label{III}

\subsection{Hierarchical Information Levels}
\begin{figure*}[!t]
    \centering
    \includegraphics[width=0.92\linewidth]{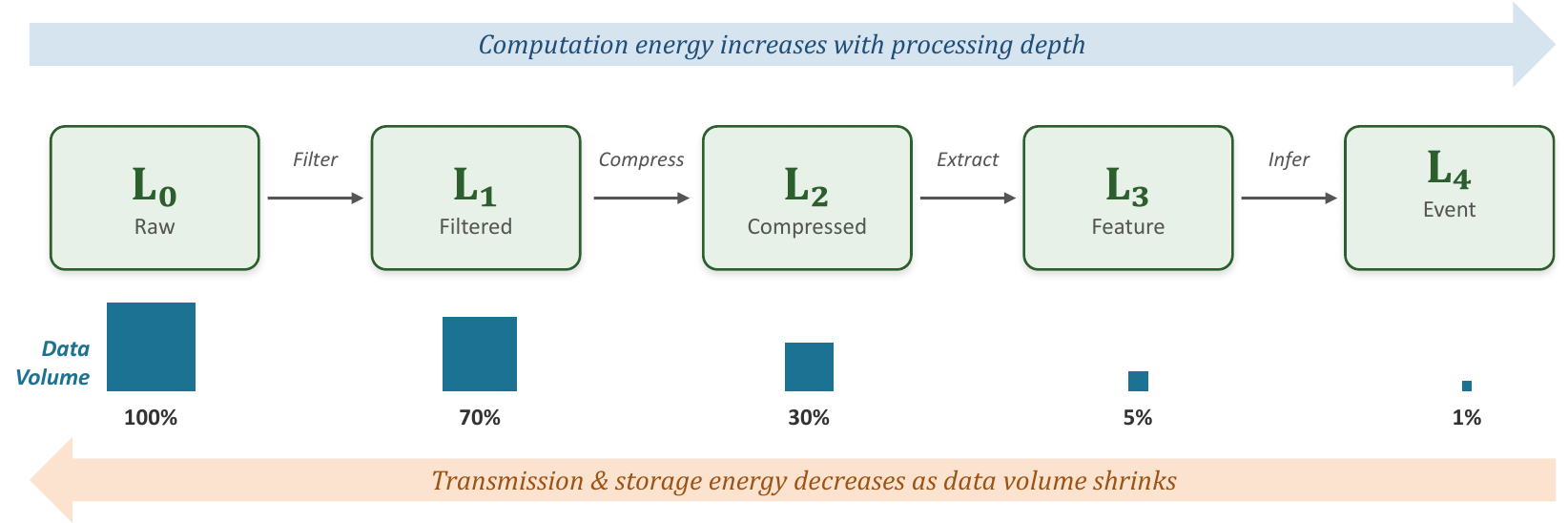}
    \caption{Multi-level information processing: raw data ($L_0$) are progressively
refined into event-level information ($L_4$), reducing data volume\textsuperscript{1}
at the cost of additional computation.%
{\footnotesize\\\textsuperscript{1}The data fractions are representative values assumed for illustration; the actual ratios depend on the data type and processing algorithms used.}}
    \label{fig:mlip}
    \vspace{-4mm}
\end{figure*}

As shown in Fig.~\ref{fig:mlip}, the refinement of data is represented using a sequence of information levels $L_0 \rightarrow L_N$. Specifically, raw data generated by USNs constitute the initial level $L_0$. Each subsequent level $L_n$ ($n \geq 1$) is generated through operations such as filtering, compression, feature extraction, fusion, or inference \cite{ref-semcom, ref-taskaqua}. These processing steps progressively reduce the data volume as information ascends toward the highest level $L_N$, where $N$ represents the total number of processing levels and varies by application.

Advancing the information from level $L_n$ to level $L_{n+1}$ requires additional computation energy, but the resulting decrease in data volume lowers the transmission energy required thereafter.
The feasible depth of refinement at any layer is determined by balancing the local processing cost against the savings in transmission and storage.
By managing this trade-off, the network distributes processing across layers in an energy-aware manner and prolongs the lifetime of energy-constrained maritime IoT networks.

{\color{black}
In practice, the processing depth can be selected using offline stage profiling and online link measurements. Before deployment, each transition from $L_n$ to $L_{n+1}$ is profiled in terms of its incremental computation and storage energy, processing delay, achieved data-reduction ratio, and information-fidelity impact. At runtime, the node or an upstream controller combines these profiles with the current per-bit link energy, session overhead, queue state, and residual battery level. It first excludes processing levels that violate latency, fidelity, memory, or energy constraints and then selects the feasible level with the lowest predicted total computation, storage, and downstream transmission cost. For resource-constrained nodes, this procedure can be implemented as a lookup table.
}

\subsection{Computation Energy}

{\color{black}
Computational energy comprises the energy consumed by arithmetic operations and by data movement across the memory hierarchy. Its total cost is governed primarily by the operation count and the volume of data transferred. Because accesses to distant memory levels involve longer interconnects and larger capacitive loads, transfers between main memory and cache can consume substantially more energy than transfers between cache and registers \cite{ref-horowitz}.

In maritime IoT networks, computation energy can be reduced by avoiding operations and memory accesses that do not contribute to the required information level. Event-triggered processing suppresses unnecessary computation when measurements remain stable, while sparse computation eliminates redundant arithmetic operations. Lightweight buffering and in-place processing further improve data locality and reduce costly main-memory accesses.}

\subsection{Transmission Energy}

{\color{black}
Transmission energy includes the energy consumed during signal transmission and reception, as well as the energy required for link establishment and maintenance. In addition to payload delivery, synchronization, idle listening, handshakes, acknowledgments, and retransmissions can introduce substantial overhead, particularly over energy-intensive underwater acoustic links \cite{ref-wakeup}.

The total transmission energy is governed primarily by the propagation medium, carrier type, and transmitted data volume. Acoustic, radio, and optical links exhibit different attenuation characteristics, achievable data rates, and per-bit energy costs across water, atmosphere, and free space \cite{ref-uwsn}. A larger data volume also keeps the transceiver active for longer, thereby increasing the cumulative transmission energy.

Transmission energy can therefore be reduced by lowering both the per-bit energy cost and the amount of data transmitted. Adaptive power control and link selection allow the transmission parameters to follow channel and visibility conditions; for example, radio links provide robust connectivity, whereas optical links offer higher data rates under favorable conditions \cite{ref-fso}. Within MLIP, refining data before transmission reduces the payload volume and transceiver active time. The benefit of further refinement consequently depends on whether the resulting transmission-energy savings compensate for the additional local computation cost.
}

\subsection{Storage Energy}

{\color{black}
Storage energy comprises the energy required to write and retain data. Random access memory (RAM) supports low-energy and high-speed read/write operations but requires continuous power to preserve its contents. In contrast, read-only memory (ROM) retains data without continuous power but incurs a higher write-energy cost \cite{ref-storage}.

Across the MLNA tiers, USNs, surface buoys, and HAPS primarily use RAM for real-time buffering and processing, while ROM preserves data during network outages and other exceptional conditions. At BSs and satellites, processed data are buffered in RAM before being transferred to ROM for long-term storage.

Within MLIP, storage energy can be reduced by decreasing both the volume of data written and the memory capacity required for buffering. Refining raw measurements into compact representations reduces RAM and ROM write operations, while application-specific RAM sizing avoids unnecessary retention energy. The resulting storage savings therefore contribute to the selection of an appropriate information-processing depth across the network layers.}

\section{Role of MLNA-MLIP in SAGSIN}

In SAGSIN, MLNA--MLIP enables maritime data to be delivered along different transmission paths while being progressively processed and refined during transmission, thereby reducing latency, enhancing data privacy and security, improving service reliability, and enabling effective energy management to prolong network lifetime, as summarized in Table~\ref{tab:roles}.

\begin{table*}[!t]
\centering
\caption{How MLNA--MLIP delivers its four benefits in SAGSIN.}
\label{tab:roles}
\renewcommand{\arraystretch}{1.3}
\begin{tabular}{p{2.4cm} p{6.2cm} p{6.2cm}}
\hline
\textbf{Benefit} & \textbf{Enabling mechanism in MLNA--MLIP} & \textbf{Representative maritime scenario} \\
\hline
Latency reduction &
In-layer processing cuts queueing delay, and refined data shorten transmission over bandwidth-limited acoustic links &
Offshore seismic monitoring: only event-level indicators forwarded for fast response \\

Data privacy &
Refinement removes raw semantic content and shortens the on-air duration, reducing the temporal attack surface &
Resource exploration: only abstract features leave the site, hiding location and infrastructure \\

Service reliability &
Intermediate buffering/refinement plus RF--optical mode switching and multi-layer path diversity &
Storm monitoring: buoys buffer locally and fall back to RF when optical links drop \\

Energy management &
Joint optimization of computation vs.\ transmission energy along the path; event-triggered reporting &
Oil-and-gas monitoring: local anomaly detection suppresses costly acoustic wake-ups \\
\hline
\end{tabular}
\end{table*}

\subsection{Latency Reduction}

Latency is a critical performance metric in maritime IoT networks.
Underwater acoustic communication operates with extremely limited bandwidth, typically on the order of kbps, making the transmission of raw sensing data highly time-consuming.
Moreover, in centralized processing architectures, large volumes of raw data introduce substantial queueing delays.
By enabling information processing to be performed at multiple network layers, the MLNA--MLIP framework reduces queueing delays caused by centralized processing.
In addition, by refining raw data into compact representations, the amount of information that needs to be transmitted is significantly reduced, which shortens transmission latency over bandwidth-constrained links, particularly underwater acoustic channels.
Distributed processing also introduces additional delay, which varies with node capabilities across layers and can be managed by adaptively adjusting the processing depth based on network conditions.

For instance, in offshore seismic monitoring, USNs continuously collect vibration and acoustic waveform data.
If raw measurements are transmitted over tens of kilometers through bandwidth-constrained acoustic links before centralized analysis, the response delay will reach several seconds due to slow transmission and queueing.
Under the MLNA--MLIP framework, preliminary feature extraction and anomaly detection can be performed at source or intermediate nodes, and only event-level indicators are forwarded upstream.
This significantly shortens the critical response time while still enabling processed summaries to be delivered for centralized aggregation and long-term assessment.

\subsection{Data Privacy}

Data privacy and security are critical concerns in maritime IoT networks.
In SAGSIN, maritime data are transmitted over long distances and multiple network layers through wireless communication media, which increases their exposure to eavesdropping and interception during transmission.
The MLNA--MLIP framework enhances data privacy and security by reducing the exposure of sensitive raw data through information processing.
Even if the processed packets are intercepted by adversaries, the information remains unintelligible without application-specific context, which significantly limits information leakage.
In addition, by refining data, MLNA--MLIP shortens the transmission duration of wireless signals, thereby reducing the temporal attack surface.
As a result, the time window during which an adversary can detect and localize transmitted signals is significantly compressed, making it more difficult to exploit or jam maritime communications.

For example, in offshore marine resource exploration, raw sensing data such as high-resolution imagery, acoustic waveforms, and structural vibration signals contain sensitive operational information, including exploration locations and infrastructure conditions.
Transmitting such sensitive data over open wireless links increases the risk of information leakage.
Under the MLNA--MLIP framework, only extracted features are forwarded across long-distance links, which significantly reduces data volume and shortens transmission time.
Even if intercepted, these abstract representations lack sufficient semantic content to reconstruct the original operational context, thereby reducing privacy risks without eliminating centralized data aggregation.

\subsection{Service Reliability}

Service reliability is a critical requirement in maritime IoT networks, where harsh maritime environments result in unstable communication links and intermittent connectivity.
By allowing data to be processed and refined at intermediate nodes, the MLNA--MLIP framework reduces the reliance on high-quality links at every transmission hop.
When temporary link impairments occur, intermediate nodes can buffer, refine, or temporarily store information and forward compact representations once connectivity is restored, thereby improving the reliability of data delivery.
Moreover, MLNA--MLIP leverages multiple transmission modes to improve service reliability. In particular, RF links provide robust connectivity with limited capacity, whereas optical links provide high-capacity transmission when visibility is favorable.
In addition, path diversity is exploited by enabling data to be forwarded over alternative transmission paths across different network layers in SAGSIN, so that link failures along a single path do not disrupt reliable data delivery.

As an example, consider offshore environmental monitoring during severe weather conditions such as storms.
Strong winds and dense clouds temporarily disrupt optical links between surface buoys and aerial platforms.
With the MLNA--MLIP framework, buoys can buffer data locally and switch to more robust RF transmission modes when optical communication becomes unavailable.
Meanwhile, alternative transmission paths across different network layers can be utilized to bypass impaired links.

\subsection{Energy Management}

Energy is the most critical constraint in maritime IoT networks.
USNs are typically battery-powered and deployed in harsh underwater environments, where maintenance and recharging are extremely difficult.
Network lifetime is commonly defined as the time when 20\% of the USNs fail due to energy depletion.
In addition, surface buoys and aerial HAPS are also equipped with batteries, but their long-term operation relies on intermittent energy harvesting from renewable sources, including solar and wave energy, whose availability depends on environmental conditions.
Moreover, in maritime IoT networks, data collected by USNs experience heterogeneous transmission paths depending on their distances from the coastline.
Specifically, maritime data traverse different transmission modes, such as acoustic channels underwater and RF/optical links in the air, each with distinct per-bit energy costs.
While information processing reduces transmission load by refining data, it introduces additional computation energy.
The MLNA--MLIP framework balances computation and transmission energy by jointly optimizing data processing and transmission decisions along the path.

For example, in long-term offshore oil and gas monitoring deployments, USNs are required to collect pressure and vibration measurements for several months without maintenance.
The MLNA--MLIP framework manages the trade-off between computation and transmission energy.
Lightweight anomaly detection and data refinement are performed locally at USNs, so that only abnormal summaries are reported upstream.
By suppressing redundant transmissions and reducing the frequency of acoustic link activation, the number of costly wake-up and handshake operations is significantly decreased, thereby extending node lifetime.

\section{Case Study: Offshore Monitoring}

To make the MLNA--MLIP trade-offs concrete, we consider a representative offshore oil-and-gas monitoring deployment. A grid of fixed USNs continuously samples pressure and vibration waveforms on the seabed and must deliver the resulting information to the core network through a surface buoy and the upper SAGSIN tiers. We compare two strategies. In the \emph{baseline} strategy, each USN transmits its raw measurements ($L_0$) to the buoy in real time over the acoustic link. 
{\color{black} In the \emph{MLNA--MLIP} strategy, each USN refines its measurements locally---filtering, compressing, extracting features, and finally distilling them into event-level summaries ($L_4$)---and transmits only when an event is detected. Intermediate levels ($L_1$, $L_2$, $L_3$) correspond to partial refinement. 
The data fractions assigned to these levels should be obtained by profiling the sensing modality and processing pipeline, including the achievable compression ratio, event occurrence rate, and required information fidelity \cite{ref-compress}. Accordingly, the values adopted in this case study are intended to reveal the qualitative computation--transmission trade-off, whereas these scenario-dependent fractions should be measured using the target sensor streams and processing algorithms in practical deployments.}
The representative parameters, consistent with the physical figures used throughout the article, are summarized in Table~\ref{tab:params}.

\begin{table*}[!t]
\centering
\caption{Representative case-study parameters (illustrative).}
\label{tab:params}
\renewcommand{\arraystretch}{1.15}
\begin{tabular}{p{6cm} p{8cm}}
\hline
\textbf{Parameter} & \textbf{Value} \\
\hline
USN battery capacity & 200 Wh ($\approx$ 0.72 MJ) \\
Raw data volume $L_0$ & 50 Mbit/day \\
Data fraction $L_1/L_2/L_3/L_4$ & 70\% / 30\% / 5\% / 1\% of $L_0$ \cite{ref-compress}\\
Acoustic energy per bit @ 5 km & $\sim$ 2 mJ/bit (10 W, 5 kbps) \cite{ref-wakeup} \\
Handshake / wake-up energy & $\sim$20 J per session \cite{ref-wakeup}  \\
On-node processing energy ($L_0\!\to\!L_n$) & depth-dependent, from $\sim$0 ($L_1$) up to the order of the transmission energy for deep inference ($L_4$) \\
RAM / ROM write energy & 0.3--1.2 / 100--850 nJ/bit \cite{ref-storage}\\
Network lifetime criterion & 20\% of USNs depleted \cite{ref-wakeup}\\
\hline
\end{tabular}
\vspace{-3mm}
\end{table*}

\textbf{Simulation setup.} 
{\color{black}We model each USN over a 24-hour cycle with three energy
terms: computation, transmission, and storage. Transmission energy comprises a fixed
per-session cost for wake-up, synchronization, and handshaking, plus a per-bit cost
given by the acoustic energy per bit times the number of bits sent. 
The projected node lifetime is then calculated by dividing
the battery capacity by this daily consumption, assuming that the
representative workload repeats.}
Following the
strong attenuation of underwater acoustic links, the per-bit energy is taken to grow
with the square of the USN-to-buoy distance, normalized to the 5 km value in Table~\ref{tab:params}.
Computation energy accounts for the local refinement from $L_0$ to the target level,
and grows with processing depth as heavier feature extraction and inference are
performed; storage energy follows the per-bit RAM/ROM write costs. The daily energy
consumption of a node is the sum of these terms. The node lifetime is its battery capacity divided
by its daily energy consumption, and the network lifetime is reached when 20\% of the USNs are
depleted. All parameters take the representative values listed in Table~\ref{tab:params} and are intended to
expose the qualitative trade-offs rather than to predict absolute lifetimes.

\begin{figure}[!t]
    \centering
    \includegraphics[width=0.92\linewidth]{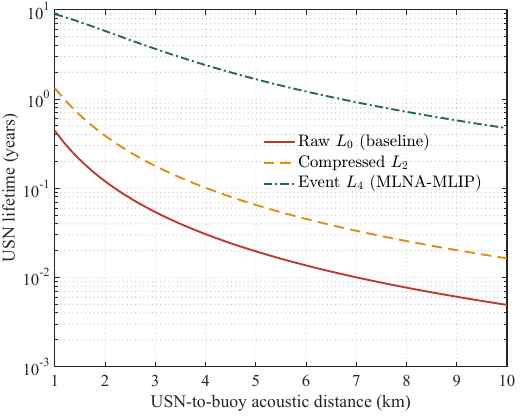}
    \caption{USN lifetime versus acoustic distance for different processing depths. Deeper refinement extends lifetime, and the advantage grows with distance.}
    \label{fig:lifetime}
    \vspace{-2mm}
\end{figure}

\textbf{Lifetime versus distance.} Fig.~\ref{fig:lifetime} shows the resulting USN lifetime as a function of the USN-to-buoy acoustic distance for three processing depths. Under the baseline strategy, transmission energy dominates: delivering the full $L_0$ stream over a long acoustic link drains the battery within weeks, and the lifetime collapses as distance grows. Refining to the compressed level $L_2$ already extends the lifetime several-fold by cutting the transmitted volume. Refining all the way to the event level $L_4$ yields the longest lifetime, because both the transmitted volume and the number of acoustic sessions are sharply reduced; at this depth the residual energy consumption is dominated by the fixed handshake cost rather than by data transmission. 
{\color{black}
This indicates that the benefit of refinement increases with transmission distance and is particularly pronounced over underwater acoustic links, where the per-bit transmission cost is high.
}

\begin{figure}[!t]
    \centering
    \includegraphics[width=0.92\linewidth]{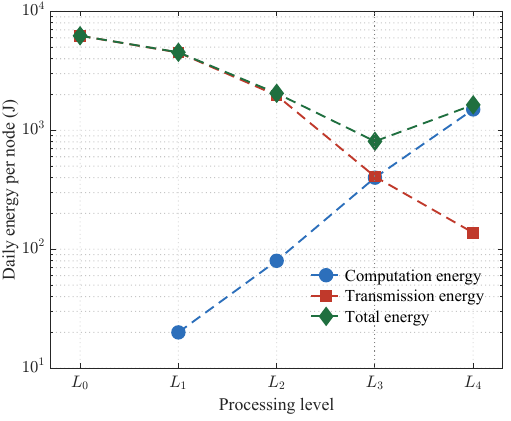}
    \caption{Energy trade-off across processing levels. The total energy is minimized at an optimal depth $N^*$ that balances computation against transmission.}
    \label{fig:tradeoff}
    \vspace{-2mm}
\end{figure}

\textbf{The optimal processing depth.} Refining data is not free, however. Fig.~\ref{fig:tradeoff} decomposes the daily energy into its computation and transmission components as a function of the processing level, for a node at a moderate distance where the acoustic link is comparatively cheap. As the processing level deepens, the transmission energy falls monotonically, but the computation energy rises, steeply once the node performs heavy on-board inference. The total energy is therefore U-shaped, reaching a minimum at an intermediate \emph{optimal depth} $N^*$: beyond this point the cost of the next refinement step exceeds the transmission energy it saves. 
The location of $N^*$ is not fixed. For a USN on a long, expensive acoustic link, the optimum shifts toward the deepest level $L_4$; for the moderate, comparatively cheap link considered in Fig.~\ref{fig:tradeoff} it falls at $L_3$; and for a surface buoy or HAPS communicating over an inexpensive RF or optical link it shifts toward still shallower levels, since further on-node processing would waste more energy than it saves. 
{\color{black}
This is exactly the cross-layer behavior anticipated in Sec.~\ref{III}: each tier should refine data only to the depth at which the transmission saving still compensates for the additional processing cost.
}

\textbf{Real-time versus accumulated reporting.} Finally, the reporting policy interacts with energy. Transmitting every measurement in real time minimizes latency but multiplies the number of acoustic sessions, and thus the handshake energy. Accumulating measurements and reporting them in batches reduces the session count and the associated overhead, at the cost of higher latency. MLNA--MLIP reconciles these objectives through event-triggered reporting: under stable conditions the node accumulates and refines data, suppressing redundant transmissions, while anomalies are reported immediately. The framework thus preserves the timeliness of critical events and the integrity of long-term records without paying the handshake cost of continuous real-time streaming.

\section{Open Challenges and Future Directions}
While MLNA--MLIP offers a coherent design lens, several problems must be solved before it can be deployed at scale.

{\color{black} \emph{Adaptive depth selection.} As the case study shows, the optimal processing depth $N^*$ shifts with link cost, channel state, energy reserves, and traffic. Hand-tuned, per-tier rules cannot track these dynamics, yet underwater nodes lack the compute and energy budget for heavy online optimization. Lightweight learning-based controllers that select the processing depth based on locally observable information, such as channel quality, residual battery level, and data similarity, are promising candidates, but their robustness in heterogeneous SAGSIN environments requires further investigation.}

\emph{Cross-layer link--computation co-scheduling.} Acoustic, RF, and optical links differ by orders of magnitude in rate and per-bit energy, and their availability fluctuates with weather and mobility. Decisions about \emph{where} to process and \emph{which} link to use must therefore be made jointly rather than layer by layer, under intermittent energy harvesting and time-varying topology. Distributed schedulers that remain stable under such uncertainty, without global state, are still missing.

{\color{black}
\emph{Complexity-aware orchestration and failure handling.} Jointly optimizing processing, routing, buffering, and link selection across four tiers can impose prohibitive control overhead, while rapidly changing channel and energy states may render centralized decisions stale. Hierarchical orchestration is therefore needed, allowing local nodes to make lightweight refinement and fallback decisions while coordinators update cross-tier policies. Key challenges are to bound signaling overhead, isolate failures, and maintain basic service when an aerial, satellite, or optical link becomes unavailable.}

\emph{Semantic and task-oriented refinement.} Moving from generic compression toward task-aware representations could push transmitted volumes far below the fractions assumed here. The open question is how to guarantee that event- and feature-level data retain sufficient fidelity for downstream analytics, and how to bound the reconstruction error when raw measurements are never recovered.

{\color{black}\emph{Standardization, security, and validation.} Interoperable interfaces across the four tiers are a prerequisite for multi-vendor SAGSIN deployments, as are lightweight security mechanisms for refined representations exchanged over exposed maritime links \cite{std-itu-y3226}. Finally, the trade-offs quantified here rest on illustrative parameters; field validation with real modems and workloads is needed to confirm the predicted gains.}

\section{Conclusion}

This article introduced MLNA--MLIP, an edge-computing paradigm that couples a four-tier SAGSIN architecture with a hierarchy of information levels for maritime IoT. By characterizing computation, transmission, and storage energy across tiers, we showed that progressively refining data along the underwater-to-space path trades a modest local processing cost for substantial savings in transmission and storage, simultaneously reducing latency, limiting the exposure of sensitive raw data, improving delivery reliability through buffering and path/mode diversity, and extending the lifetime of energy-constrained nodes. A case study on offshore monitoring quantified these gains and revealed an optimal processing depth that minimizes total energy and shifts with the tier and link in use. 

\bibliographystyle{IEEEtran}
\bibliography{references}  

\end{document}